\documentclass[pra,aps,twocolumn,superscriptaddress,showpacs]{revtex4}
\usepackage{amsmath,amsfonts,amssymb}
\usepackage{graphicx}
\usepackage{epsfig}

\begin{document}
%**************************************************************
\title{Many-Body Effects on Nonadiabatic Feshbach Conversion in Bosonic Systems}
\author{ Jie Liu}
\affiliation{Center for Applied Physics and Technology, Peking
University, 100084, Beijing, P.R.China} \affiliation{Institute of
Applied Physics and Computational Mathematics, Beijing 100088, P. R.
China}
\author{ Bin Liu}
\affiliation{Graduate School, China Academy of Engineering Physics,
Beijing 100088, P. R. China } \affiliation{College of Physics and
Information Engineering, Hebei Normal University, 050016
Shijiazhuang, China}
\author{ Li-Bin Fu}
\affiliation{Institute of Applied Physics and Computational
Mathematics, Beijing 100088, P. R. China}
\begin{abstract}
We investigate the dynamics of converting cold bosonic atoms to
molecules when an  external magnetic field is swept across a
Feshbach resonance.  Our analysis relies on a quantum microscopic
model that accounts for many-body effects in the association
process. We show that the picture of two-body molecular production
depicted  by  the Landau-Zener model is significantly altered due to
many-body effects. In the nonadiabatic regime, we derive an analytic
expression for molecular conversion efficiency that explains the
discrepancy between the prediction of the Landau-Zener formula and
the  experimental data[Hodby et al., Phys. Rev. Lett. {\bf 94},
120402 (2005)]. Our theory is further extended to the formation of
heteronuclear diatomic molecules and gives some interesting
predictions.
\end{abstract}
 \maketitle

%**************************************************************
\section{Introduction}
The production of ultracold diatomic molecules is an exciting area
of research with important applications ranging from the search for
the permanent electric dipole moment\cite{dipole} to  BCS-BEC
(Bose-Einstein condensate) crossover physics\cite{bcs}. A widely
used production technique involves the association of ultracold
atoms into very weakly bound diatomic molecules by applying a time
varying magnetic field in the vicinity of a Feshbach
resonance\cite{Timmermans,rmp}. The underlying conversion dynamics
are usually described by the Landau-Zener (LZ) model\cite{lzf}. In
this model, the Feshbach molecular production is discussed under a
two-body configuration where a single pair of atoms is converted to
a molecule at an avoided-crossing between  atomic energy level and
molecular energy level while the molecular energy is lifted by an
applied linearly sweeping magnetic field. Thus, the molecular
production efficiency is expected to be  an exponential Landau-Zener
type\cite{mies,jpb}. However, recent experimental data on $^{85}$Rb
by the JILA group showed a large discrepancy from the Landau-Zener
formula: The value of the LZ parameter extracted from the data is 8
times larger than the prediction of LZ theory\cite{data}. The
experiment was performed under unusually low densities of the atom
cloud ( $\sim 10^{11} /cm^3$) and the data was measured in the
nonadiabatic regime so that the inverse ramp rate was less than
100$\mu s/G$. Therefore,  two- and three-body atomic decay and
collisional molecular decay rates are negligible and do not affect
the measurement. This puzzle remains unresolved and challenges our
knowledge of the big issue of  Feshbach molecular formation.

In this paper, using a many-body two-channel microscopic
Hamiltonian, we investigate the dynamics of  Feshbach molecular
formation in bosonic systems such as $^{85}$Rb. We show that
many-body effects alter the LZ picture of two-body molecular
production through dramatically distorting the energy levels near
the Feshbach resonance. With the help of a mean-field classical
Hamiltonian, we derive an analytic expression for the conversion
efficiency in the nonadiabatic regime. Our theory agrees with
experimental data. Our theory thus is extended to the Feshbach
formation of heteronuclear diatomic molecules such as
$^{85}$Rb-$^{87}$Rb and predicts that many-body effects are more
significant there.

Our paper is organized as follows. In Sec.II, we present our model.
In Sec.III, many-body effects on the conversion dynamics are
addressed and an analytic expression for conversion efficiency is
derived. In Sec.IV, we apply  our theory to explain the experimental
data. In Sec.V, we extend our theory to the heteronuclear molecules.
The final section is our conclusion.

%*****************************************************************************************
%*****************************************************************************************
\section{Model}
Considering the experimental condition that the densities of the
atom cloud is unusually low and the two- and three-body atomic decay
and collisional molecular decay rates are negligible, we exploit the
following two-channel model to describe the dynamics of converting
atoms to molecules in the bosonic system,
\begin{eqnarray}
\hat{H}=  \epsilon_{a}\hat{a}^{\dagger}\hat{a}+
\epsilon_b(t)\hat{b}^{\dagger}\hat{b}
+\frac{g}{\sqrt{V}}\left(\hat{a}^{\dagger}\hat{a}
^{\dagger}\hat{b}+\hat{b}^{\dagger}\hat{a} \hat{a}\right),
\label{hambb}
\end{eqnarray}
where $\hat{a}$ ($\hat{a}^+$) and $\hat{b}$($\hat{b}^+$) are Bose
annihilation (creation)  operators of atoms and
molecules, respectively. $g=\sqrt{4\pi\hbar^2a_{bg}\Delta
B\mu_{co}/m}$ is the atom-molecule coupling due to the Feshbach
resonance, $m$ is the mass of a bosonic atom, $a_{bg}$ is background
scattering length, $\Delta B$ is the width of the resonance, and
$\mu_{co}$ is the difference in the magnetic moment between the
closed channel and open channel state. Here, we introduce parameter
$V$ to denote the volume of trapped particles and therefore $n=N/V$
is the mean density of initial bosonic atoms. The external magnetic
field is linearly swept $B(t)=\dot{B} t$ and  crosses the Feshbach
resonance at $B_0$. The molecular energy under the field is
$\epsilon_b(t)=\mu_{co}\left(B(t)-B_0\right)$. The total number of
particles $N=\hat{a}^{\dagger}\hat{a}+2\hat{b}^{\dagger}\hat{b}$ is
a conserved constant.
\begin{figure}[t]
  \centering
     \includegraphics[width=0.45\textwidth]{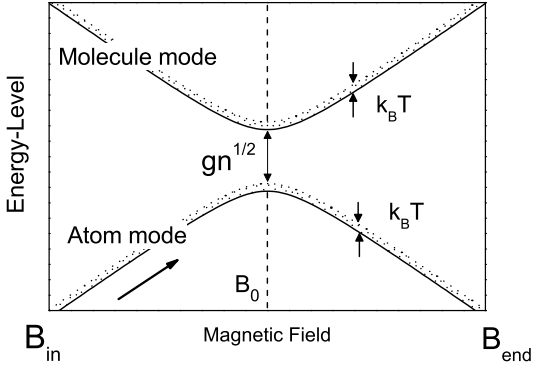}
  \caption{Schematic plot of energy levels of atomic mode and molecular mode.
  Our single-mode approximation is valid only when the energy
  distribution of the thermal particles is much smaller than
   the effective width of the Feshbach resonance, i.e.,
    $k_B T <<\sqrt{g^2 n}$. Moreover, it is required that
    the time  for the external magnetic field to sweep across the Feshbach
    resonance is smaller than the "dephasing" time
    $\tau_d = \frac{2\pi\hbar}{k_BT}$.
    For a detailed discussion  refer to  the  text.}
  \label{sche}
\end{figure}

The above single-mode model is an approximate description of the
clouds of noncondensed atoms and molecules, and is only valid when
the energy distribution of the thermal particles (characterized by
$k_B T$, $k_B$ is the Boltzman constant and $T$ is the temperature)
is much smaller than the effective Feshbach resonance width
$g\sqrt{n}$. In such cases, each 'energy band' of the thermal
particles can be approximately denoted by one energy level, as
schematically plotted by Fig.\ref{sche}. Initially, the particles on
one  level have a definite phase and the phase difference between
two levels is well defined. However, as the magnetic field sweeps
across the Feshbach resonance from $B_{in}$ to $B_{end}$ at a rate
of $\dot B$, particles will acquire additional phases that are
proportional to their individual energy and sweeping time
$(B_{end}-B_{in})/\dot B$. The varied particles in one level could
acquire different phases because they have different energies. The
validity of our single-mode approximation requires that the above
mismatch in phase or 'dephasing'  is at least smaller than $2\pi$,
which  defines a 'dephasing time' $\tau_d =
\frac{2\pi\hbar}{k_BT}$\cite{dephase}. When the time taken by the
external magnetic field to sweep across the Feshbach resonance is
smaller than the above  'dephasing' time, i.e.,
$(B_{end}-B_{in})/\dot B < \frac{2\pi\hbar}{k_BT}$, the above
dephasing effects can be ignored. The above analysis sets up a lower
bound on the sweeping rate.  So, in  the  following discussion, we focus
on the fast-swept or nonadiabatic regime in which the above
condition is satisfied.

\begin{figure}[t]
  \centering
     \includegraphics[width=0.45\textwidth]{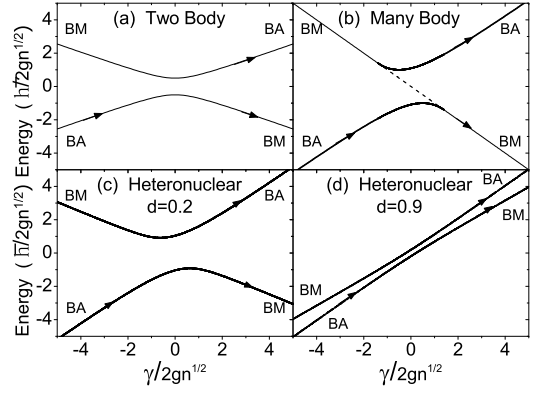}
  \caption{Energy levels versus the scaled external magnetic fields.
  a) Homonuclear two-body case;
  b) Homonuclear many-body case ($N=\infty$), the dashed line represents an additional unstable eigenstate;
  c-d) Heteronuclear many-body cases ($N=\infty$), where parameter $d$ denotes
  initial population imbalance
   between two species. }
  \label{energy}
\end{figure}

Using  the Fock states as a basis, the  Schr\"{o}dinger equation
is written as
\begin{eqnarray}
i\frac{d}{dt}|\psi\rangle=\hat{H}|\psi\rangle,
\label{sch}
\end{eqnarray}
where $|\psi\rangle=\sum_{j=0}^{N/2}c_j|2j,N/2-j\rangle$,
$|2j,N/2-j\rangle=\frac{1}{\sqrt{(2j)!(N/2-j)!}}
\left(\hat{a}^{\dagger}\hat{a}^{\dagger}\right)^{j}
\left(\hat{b}^{\dagger}\right)^{N/2-j}|0\rangle \quad (j=0,...,N/2)$
are Fock states, and  $c_j$ is the probability amplitudes on the
corresponding Fock state, respectively. The normalization condition
is  $\sum_j |c_j|^2=1$.

For the simplest case of  $N=2$, the above  Schr\"{o}dinger equation
reduces to the following two-level system of  Landau-Zener type,
\begin{eqnarray}
i \hbar \frac{d}{dt} \left(
\begin{array}{c}
c_0 \\
c_1
\end{array}
\right)=
\left(
\begin{array}{ll}
\epsilon_{b} & v/2 \\
v/2 & 2\epsilon_{a}
\end{array}
\right)
\left(
\begin{array}{c}
c_0 \\
c_1
\end{array}
\right).
\label{lzh}
\end{eqnarray}
Where  $|c_0|^2$ and $|c_1|^2$ denote the population of molecules
and atoms, respectively. For the two-level system, the energy bias
between two levels is
$\gamma=\left(2\epsilon_{a}-\epsilon_{b}\right)$  and the coupling
strength is given by $v=2g\sqrt{n}$. Initially, all particles
populate in the lower level of the atomic state, i.e.,
$c_0=0,c_1=1$. When the external magnetic field is linearly swept
across the Feshbach resonance at $\gamma \simeq 0$, a fraction of
atoms will be converted to molecules at the avoided-crossing of
energy levels. The conversion efficiency as a function of the
sweeping rate (i.e., $\dot \gamma =\mu_{co} \dot B$) and coupling
strength, takes the form\cite{lzf},
\begin{eqnarray}
\Gamma_{lz} = 1-\exp(-\frac{\pi v^2}{2\hbar\dot\gamma})=
1-\exp\left(-\frac{8 \pi^2 n\hbar |a_{bg}\Delta B|}{m
|\dot{B}|}\right).
\label{gamlz}
\end{eqnarray}
The above is  the two-body molecular production picture and  is
consistent with the result  from the coupled-channel scattering
calculation in Ref.\cite{jpb}.

Mathematically, ignoring  a total phase, the dynamics of
Eq.(\ref{lzh}) are equivalent to the following simple  classical
Hamiltonian\cite{lzt,clasicalh},
\begin{eqnarray}
\mathcal{H}_{lz}=-\gamma/\hbar  s +v/\hbar\sqrt{1-s^2}\cos\theta.
\label{lzch}
\end{eqnarray}
Where the canonical conjugate variables are the population
difference $s=|c_0|^2-|c_1|^2$  and the relative phase $\theta=\arg
c_0-\arg c_1$. The dynamics are governed by the canonical equations
of $\dot \theta = \frac{\partial \mathcal{H}_{lz}}{\partial s},\dot
s = -\frac{\partial \mathcal{H}_{lz}}{\partial \theta}$. The fixed
points satisfying $\dot s =0, \dot \theta =0$ correspond to the
extremum of system energy. These classical fixed points correspond
to the eigenstates of quantum equations (\ref{lzh}) and their
energies (corresponding to the eigenvalues of quantum eigenstates)
are calculated and plotted against the energy bias parameter
$\gamma$ in Fig.\ref{energy}a. It exhibits a typical LZ
avoided-crossing configuration. Initially,  all particles populate
in the atomic state of  $s_0 = -1$ at the left end of the lower
level. When the external field passes through the Feshbach resonance
of width $v/\hbar$ at $\gamma =0$, a fraction of atoms are converted
to molecules at the right end of the lower level, leading to a
variation in the population variable, i.e.,
\begin{equation}
s_f =2\Gamma_{lz} - 1=
1-2\exp(-\frac{\pi v^2}{2\hbar\dot\gamma}).
\label{sf}
\end{equation}

As we go beyond the above  two-body treatment to  consider the
many-body effects, the structure of the energy levels will change
dramatically and the above  LZ formula of the conversion efficiency
will be altered due to  many-body effects.

%=========================================================================
\begin{figure}[t]
  \centering
     \includegraphics[width=0.45\textwidth]{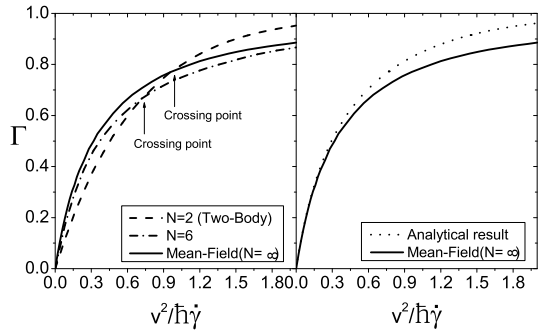}
  \caption{Feshbach molecular production efficiency versus the
   scaled inverse sweep rates. As we increase the number $N$ of the particles, the
   many-body result converges to the mean-field curve. }
  \label{mf}
\end{figure}
%---------------------------------------------------------------------------

%*****************************************************************************************
%*****************************************************************************************

\section{Many-Body Effects in Forming Homonuclear Feshbach Molecules}

To include many-body effects, we need to solve full $\frac{N}{2}+1$
dimensional quantum equations (\ref{sch}). Using the basis of Fock
states, the Schr\"odinger equation is rewritten as
\begin{equation}
i\frac{d c_j}{dt} = \sum_k H_{jk}c_k, (j,k =0,1,...,N/2)
\end{equation}
where the  Hamiltonian matrix elements are
$H_{jk}=<2j,N/2-j|H|2k,N/2-k>$. For $j=k$, $H_{jj}=j\gamma$;
for $j\neq k$, $H_{jk}=0$ except
$H_{j,j+1}=H_{j+1,j}=\sqrt{(j+1)(2j+1)(N/2-j)/2N}v$.

The above differential equations do not have  explicit analytic
solutions. We thus solve them numerically using the 4-5th order
Runge-Kutta algorithm with an adaptive time-step. Our result is
presented in Fig.\ref{mf}, which shows that the molecular conversion
is altered due to many-body effects. Interestingly, there exists a
crossing point between the two-body conversion curve and the
many-body conversion curve. As the scaled inverse sweep rate is
below this crossing point, the many-body effects enhance the
molecular conversion efficiency, while as the scaled inverse sweep
rate is above the crossing point, the many-body effects suppress the
molecular conversion efficiency. The location of the crossing point
is dependent on the total particle number $N$ and shifts to the
right as $N$ increases. For $N=6$, the crossing point corresponds to
$v^2/\hbar \dot \gamma =0.7$. It shifts to  one  as $N=\infty$.

Below, with the help of angular momentum operators, we deduce an
analytic expression for
 the atom-molecule conversion efficiency under the mean field
approximation.

The angular momentum  operators are introduced as
follows\cite{twomode},
\begin{eqnarray}
\hat{L}_{x} &=& \sqrt{2}\frac{\hat{a}^{\dagger}\hat{a}
^{\dagger}\hat{b}+\hat{b}^{\dagger}\hat{a} \hat{a}
}{N^{3/2}}, \\
\hat{L}_{y} &=& \sqrt{2}i\frac{\hat{a}^{\dagger}\hat{a}
^{\dagger}\hat{b}-\hat{b}^{\dagger}\hat{a} \hat{a}
}{N^{3/2}}, \\
\hat{L}_{z} &=&
\frac{2\hat{b}^{\dagger}\hat{b}-\hat{a}^{\dagger}\hat{a}}{N}.
\end{eqnarray}
 The operator $L_z$ denotes the atom-molecule population
imbalance, and $L_x,L_y$ describe the coherence of atoms and
molecules. They compose a generalized Bloch
representation\cite{{vardi2}}. The commutators between the operators
are $ \left[\hat{L}_z,\hat{L}_x \right]= \frac{4i}{N}\hat{L}_y,
%-----------------------------------------------
\left[\hat{L}_z,\hat{L}_y \right]=
-\frac{4i}{N}\hat{L}_x,
%-----------------------------------------------
\left[\hat{L}_x,\hat{L}_y \right] =
\frac{i}{N}\left(1-\hat{L}_z\right)\left(1+3\hat{L}_z\right)
+\frac{4i}{N^2}. \label{commu} $ $\hat{L}_x, \hat{L}_y, \hat{L}_z$
do not span SU(2) because the commutator $\left[\hat{L}_x,\hat{L}_y
\right]$  yields a quadratic polynomial in $L_z$. The generalized
Bloch surface is determined by the conserved relationship $
(\hat{L}_x)^2+(\hat{L}_y)^2 =
\frac{1}{2}\left(1+\hat{L}_z\right)\left(1-\hat{L}_z\right)^2
 +\frac{2}{N}\left(1-\hat{L}_z\right)
+\frac{4}{N^2}\hat{L}_z.$ The Hamiltonian (\ref{hambb}) becomes
$\hat{H}=-\frac{N}{4} \gamma \hat{L}_z+\frac{\sqrt{2}N}{4} v
\hat{L}_x$\cite{note}. The Heisenberg equations are $ i \hbar
\frac{\textit{d}}{\textit{d}t}\hat{L}_j =
\left[\hat{L}_j,\hat{H}\right], j=x,y,z$.

In the mean field limit where $N\rightarrow\infty$, all the above
commutators vanish. Therefore, it is appropriate to replace
 $L_x$, $L_y$
and $L_z$ by their expected values $u,w$, and $s$, respectively.
Noting the constraint $u^2+w^2=\frac{1}{2}(s-1)^2(s+1)$ and
introducing the conjugate angular variable $\theta=\arctan( w/u )$
denoting  the relative phase between atoms and molecules, the
Heisenberg equations can be replaced by a classical Hamiltonian of
the  form
\begin{eqnarray}
\mathcal{H}_m &=& -\gamma/\hbar s + v/\hbar\sqrt{(1-s^2)(1-s)}\cos
\theta, \label{clas}
\end{eqnarray}

To understand the dynamics, we first look at the fixed points
$\dot{s}= \dot{\theta}=0$. The energies for these fixed points make
up energy levels of the system, as shown in Fig.\ref{energy}b. The
structure of these energy levels changes dramatically compared to
the two-body case.  We observe: (\textit{i}) There are two fixed
points when $|\gamma/v|$ is large enough: one for the bosonic
molecule (BM) and the other for the bosonic atom (BA). (\textit{ii})
When $|\gamma/v|<\sqrt{2}$, there is an additional fixed point with
$s=1$. However, this fixed point is a saddle point corresponding to
dynamically unstable quantum states\cite{unstable}. Eq.(\ref{clas})
and the energy spectra in Fig.\ref{energy}b are the same as those
obtained for the two-mode atom-molecule Fermi
system\cite{vardi2,liu66} except that the  sign of the magnetic
field is reversed.

Compared to Hamiltonian (\ref{lzch}), the coupling term in many-body
Hamiltonian (\ref{clas}) is renormalized by a factor $\sqrt{1-s}$.
So, the Fehsbach resonance width that is proportional to the
coupling  either broadens or shrinks depending on the factor. For
the fast sweep case, $s$ should be not far from its initial
value $-1$, therefore, the resonance width broadens and we expect
that many-body effects enhance the atom-molecule conversion. In
contrast, for the slow sweep case,  $s$  may take a value close to
$1$, therefore, the resonance width shrinks.  We then  expect that
the many-body effects suppress the atom-molecule conversion compared
to the two-body Landau-Zener formula. The above analysis reveals the
mechanism behind the crossing phenomenon exhibited in Fig.\ref{mf}.

To derive an approximate analytic  expression for the conversion
efficiency, we introduce an effective coupling $v_{eff}$ as,
\begin{equation}
v_{eff}= v\sqrt{1-s^*}.
\end{equation}
Where  $s^*$ can be approximately taken as the average between
initial value $s_0=-1$ and the final value $s_f$, i.e.,
$s^*=(-1+s_f)/2$. Using the relation
$\Gamma_m=2<\hat{b}^{\dagger}\hat{b}>/N=(1+s_f)/2$ and formula (4),
we obtain a self-consistent formula for the many-body conversion
efficiency $\Gamma_m$,
\begin{eqnarray}
\Gamma_{m}\simeq 1-\exp(-\frac{\pi
v^2(2-\Gamma_m)}{2\hbar\dot\gamma}). \label{gamm}
\end{eqnarray}
The above self-consistent equation for the conversion efficiency
$\Gamma_m$ can be readily solved using  the  iteration method. The result
is presented in Fig.\ref{mf}. We also numerically solve the mean
field equations using the Runge-Kutta step-adaptive algorithm. They
are in good agreement, especially in the nonadiabatic regime of fast
sweep rates (see Fig.\ref{mf}). For the slow sweep case, the above
formula overestimates molecular conversion slightly.

%=========================================================================
\begin{figure}[t]
  \centering
     \includegraphics[width=0.45\textwidth]{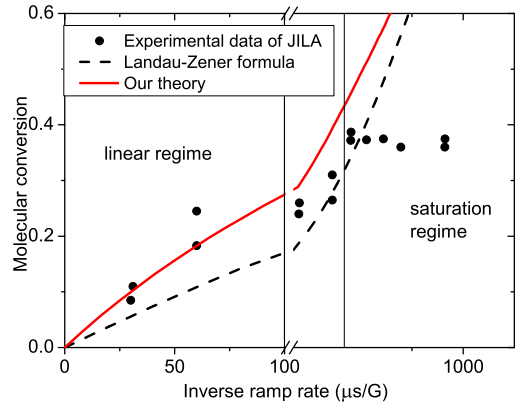}
  \caption{Molecular conversion efficiency versus the inverse sweeping rates.  }
  \label{expdata}
\end{figure}
%---------------------------------------------------------------------------

%*****************************************************************************************
%*****************************************************************************************
\section{Comparison with Experiment of $^{85}$Rb}
Now we apply our theory to $^{85}$Rb experiment by the JILA
group\cite{data}. The atoms are held in a purely magnetic "baseball"
trap. For efficient evaporation, the bias field is held at 162 G,
where the scattering length  is positive. For slow magnetic field
ramps, Rb$_2$ molecules are produced only when the field is ramped
upward through the resonance, which is located at 155G. Hence, the
first step in molecule production is to rapidly jump the magnetic
field from 162G to 147.5G. They then sweep the field back up to 162G
at a chosen linear rate, producing molecules as they pass through
the Feshbach resonance. The initial conditions of the atomic cloud
are $N=87000$ and $n = 1.3\times10^{11} cm^{-3}$. The Feshbach
resonance parameters are $a_{bg} = -443 a_0, \Delta B = 10.71G,$ and
$\mu_{co}=-2.33\mu_B,$ where $a_0$ and $\mu_B$ are the Bohr radius
and Bohr magneton, respectively. The thermal cloud of the particles
is at temperature $T=40$nK.

In the first part of their experiment, they measured the  molecular
conversion efficiency as a function of the inverse ramp rate. A
typical data set is shown in Fig.\ref{expdata}. The data are mainly
divided into three regimes, i.e., the linear increase regime where
the inverse ramp rates are less than 100 $\mu s/G$, the saturation
regime where the inverse ramp rates are larger than 200 $\mu s/G$,
and the transition regime in between. The single-mode approximation
exploited in our theory requires that the resonance width is much
larger than the energy distribution of the particles.  In this
experiment, we see that the ratio $g\sqrt{n}/k_B T$ is around $20$
at $T=40$ nK. While in the saturation regime, the molecular
conversion rates are found to saturate around $37\%$. From our
"dephasing" criterion discussed in Sec.II, i.e.,
$(B_{end}-B_{in})/\dot B < \frac{2\pi\hbar}{k_BT}$ and that $B_{in}
=147.5$G, $B_{end}=162$G, we have $1/\dot B < 82\mu s/G$. So, in our
discussion, we only focus on the first regime. In Fig.\ref{expdata},
we plot the results from our many-body theory, which show  good
agreement with the experimental data in the linear regime. As a
comparison, we also plot the result from the Landau-Zener formula,
which shows a pronounced deviation from the experimental data.
%=========================================================================
\begin{figure}[t]
  \centering
     \includegraphics[width=0.45\textwidth]{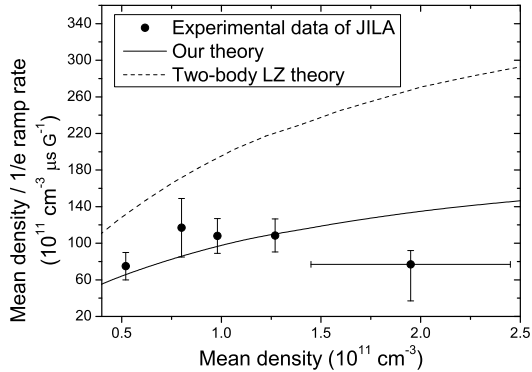}
  \caption{The ratios of mean density over 1/e ramp rate, with respect to
  mean density. Our theory shows a good
agreement with the experiment for four low density points but is
 obviously  larger than the final point. At the high density, the cloud
experienced significant heating during the ramps across the
resonance, hence the density of the final point has significant
uncertainty\cite{data}.}
  \label{den}
\end{figure}
%---------------------------------------------------------------------------

In the second part of the experiment, to compare with Landau-Zener
theory quantitatively, the JILA group measured the ratio between
mean density and $1/e$ ramp rate as a function of mean density. They
found that the Landau-Zener parameter predicted from the two-body
theory is roughly 1/8 of the value extracted from the experimental
data. They use the formula $N_{mol} = N_{max} (1-e^{-\alpha n\Delta
B a_{bg}/\dot B})$ to fit the experimental data on molecular
conversion, where $N_{max}$ is the asymptotic number of molecules
created for a very slow ramp,$\dot B$ is the magnetic field sweeping
rate, and $\alpha$ is a fitting parameter. $\delta_{LZ}=\alpha
n\Delta B a_{bg}/\dot B$ is the Landau-Zener parameter. The
saturation data in Fig.\ref{expdata} indicate that $N_{max}/N
=37\%$. The $1/e$ ramp rate $\dot B_{1/e}$ is defined as that at
$\dot B_{1/e}$, $\delta_{LZ} =1$ and $N_{mol}/N_{max} =63\%$. It was
then claimed that the data support a constant value for $n/\dot
B_{1/e}$ (see Fig.\ref{den}). The value for $\alpha$, extracted from
the experimental data, is $4.5\times 10^{-7}m^2s^{-1}$. However, the
two-body Landau-Zener formula (4) predicts $\alpha = 8\pi^2\hbar/m =
5.9\times 10^{-8}m^2s^{-1}$,
 roughly $1/8$ of the
experimental data.

Now we apply our many-body theory to resolve this puzzle. At
$B_{1/e}$, the molecular conversion efficiency is
$N_{mol}/N=37\%\times 63\%=23\%$. In the nonadiabatic regime, our
many-body formula (\ref{gamm}) is simplified as $\Gamma_{m} \simeq
\frac{16 \pi^2 n\hbar |a_{bg}\Delta B|}{m |\dot{B}|}$. Substituting
$\Gamma_m =23\%, B=B_{1/e}$ into the above formula, we have
$n/\dot{B}_{1/e} = \frac{0.23m}{16 \pi^2 \hbar |a_{bg}\Delta B|}=
105\times 10^{11} cm^{-3}\mu s G^{-1}$, which is good agreement with
the experimental data of the fourth scatter  in Fig.\ref{den}.

To compare with two-body LZ formula Eq.(4),  we see that, the
many-body effects change the $1/e$ rate in the non-adiabatic regime
by a factor of 2. The above analysis uncovers the physics behind the
$1/8$ deviation. The factor $1/8$ is the product of following three
factors: 0.37 is from the maximum conversion rate, 0.63 is from the
definition of the 1/e ramp rate, and 1/2 comes from many-body
effects. .

Our calculations are extended to the cases of varied spatial
densities. As mentioned above,  in Ref.\cite{data}, the formula
$N_{mol} = N_{max} (1-e^{-\alpha n\Delta B a_{bg}/\dot B})$ is used
to fit experimental data on molecular conversion. Accordingly, the
$1/e$ ramp rate $B_{1/e}$ corresponds to $N_{mol}/N_{max} =
1-1/e=63\%$. Our many-body theory Eq.(13) predicts that
$n/\dot{B}_{1/e} = \frac{0.63m}{16 \pi^2 \hbar |a_{bg}\Delta B|}
\frac{N_{max}}{N}$.
 Because
the maximum molecular conversion efficiency (i.e., $N_{max}/N$) is a
function of peak phase space density as revealed  in Fig.2 in
Ref.\cite{data} and the spatial density is proportional to peak
phase space density at the fixed temperature, we claim that
$n/\dot{B}_{1/e}$ is spatial density $n$ dependent through
$N_{max}/N$. The $N_{max}/N$ as a function of density is read out
from Fig.2 in Ref.\cite{data}. Thus, our theoretical curve is
plotted against the experimental data in Fig.\ref{den}. It shows
good agreement with the experiment for four low density points but
is obviously larger than the final point. At  high density, the
cloud experienced significant heating during the ramps across the
resonance, hence the density of the final point has significant
uncertainty (i.e., see the caption of Fig.1 of \cite{data}). The
result from two-body LZ theory is also presented in Fig.\ref{den}
for comparison. It is twice as large as that of many-body theory,
and obviously deviates from the experimental data.

%*****************************************************************************************
%*****************************************************************************************
\section{Many-Body Effects in Forming Heteronuclear Feshbach Molecules}

In the  above discussion, we investigated the dynamics of Feshbach
converting single atomic species  to homonulcear diatomic molecule.
Actually, the Feshbach resonance technique has been used to produce
heteronuclear molecules from two or more species of
atoms\cite{hetero}. These ultracold heteronuclear molecules in
low-lying vibrational states are of particular interest since they
could be a permanent dipole moment due to the unequal distribution
of electrons. Other proposals for using the polar molecules include
quantum computation\cite{appl1} and testing fundamental
symmetry\cite{appl2}.

In this section, we extend our discussion to the two-species atom
case and show that the heteronuclear molecular production efficiency
is more significantly altered due to many-body effects.  The
many-body three-channel Hamiltonian for the heteronuclear system
reads \cite{heter}
\begin{eqnarray}
\hat{H} &=&  \epsilon_{a1}\hat{a}_1^{\dagger}\hat{a}_1
+\epsilon_{a2}\hat{a}_2^{\dagger}\hat{a}_2
+\epsilon_b(t)\hat{b}^{\dagger}\hat{b}      \notag \\
&& +\frac{g}{\sqrt{V}}\left(\hat{a}_1^{\dagger}\hat{a}_2
^{\dagger}\hat{b}+\hat{b}^{\dagger}\hat{a}_1 \hat{a}_2\right).
\label{heter}
\end{eqnarray}
where $\hat{a}_1$ ($\hat{a}_1^{\dagger}$), $\hat{a}_2$
($\hat{a}_2^{\dagger}$) are annihilation (creation) operators of the
heteronuclears atoms and $\hat{b}$ ($\hat{b}^{\dagger}$) are
annihilation (creation) operators of molecules,
$g=\sqrt{2\pi\hbar^2a_{bg}\Delta B\mu_{co}/m'}$ is the atom-molecule
coupling strength, and $m'=m_1 m_2/(m_1+m_2)$ is the reduced mass of
two atom scattering. The total number of particles
$N=N_{a1}+N_{a2}+N_{b}=\hat{a}_1^{\dagger}\hat{a}_1+\hat{a}_2^{\dagger}\hat{a}_2+2\hat{b}^{\dagger}\hat{b}$
is a conserved constant. The density $n=N/V$.

 Using  the Fock states as a basis, the Schr\"odinger equation is
written as
\begin{eqnarray}
i\frac{d}{dt}|\psi\rangle=\hat{H}|\psi\rangle, \label{dsch}
\end{eqnarray}
where $|\psi\rangle=\sum_{j=0}^{N/2}c_j|j,j,N/2-j\rangle$,
$|j,j,N/2-j\rangle=\frac{1}{\sqrt{j!j!(N/2-j)!}}
\left(\hat{a}_1^{\dagger}\hat{a}_2^{\dagger}\right)^{j}
\left(\hat{b}^{\dagger}\right)^{N/2-j}|0\rangle \quad (j=0,...,N/2)$
are Fock states, and  $c_j$ is the probability amplitudes on the
corresponding Fock state, respectively. The normalization condition
is that $\sum_j |c_j|^2=1$.

For $N=2$,
the Schr\"{o}dinger equation
reduces to
\begin{eqnarray}
i \hbar \frac{d}{dt} \left(
\begin{array}{c}
c_0 \\
c_1
\end{array}
\right)=
\left(
\begin{array}{ll}
\epsilon_{b} & v/2\sqrt{2} \\
v/2\sqrt{2}  & \epsilon_{a1}+\epsilon_{a2}
\end{array}
\right)
\left(
\begin{array}{c}
c_0 \\
c_1
\end{array}
\right). \label{dlzh}
\end{eqnarray}
Where $|c_0|^2$ and $|c_1|^2$ denote the population of molecules and
atoms, respectively. The energy bias $\gamma=
\epsilon_{a1}+\epsilon_{a2}-\epsilon_{b}$. Then, two-body molecular
production efficiency is $ 1-\exp(-\frac{\pi
v^2}{4\hbar\dot\gamma}).$ Comparing the above expression with
Eq.(\ref{gamlz}), a $1/2$ factor emerges in the exponent. This is
due to the distinguishability  between two atomic species that
decreases the effective density of each atomic species.

For the two-species case, the number of particles in each species
may not be identical. Therefore, we introduce a parameter $d$ to
denote the population imbalance between the two species, i.e.,
$d\equiv (N_{a1}-N_{a2})/N$  assuming  that $N_{a1} > N_{a2}$.
Our concern is the conversion efficiency of type-2 atoms, i.e.,
$\Gamma_d = 2N_b/(1-d)N$ when the magnetic field is swept across the
resonance. We will show that $d$ is an important parameter in
forming the heteronuclear molecule. For the larger population
imbalance, the heteronuclear molecule production is more
significantly altered due to the many body effect.

For the heteronuclear system, the Bloch space is expanded by
following three operators $ \hat{L}_{x} =
2\sqrt{2}\frac{\hat{a}_1^{\dagger}\hat{a}_2
^{\dagger}\hat{b}+\hat{b}^{\dagger}\hat{a}_1 \hat{a}_2 }{N^{3/2}},
\hat{L}_{y} = 2\sqrt{2}i\frac{\hat{a}_1^{\dagger}\hat{a}_2
^{\dagger}\hat{b}-\hat{b}^{\dagger}\hat{a}_1 \hat{a}_2 }{N^{3/2}},$
and $ \hat{L}_{z} =
\frac{2\hat{b}^{\dagger}\hat{b}-\hat{a}_1^{\dagger}\hat{a}_1
-\hat{a}_2^{\dagger}\hat{a}_2}{N}. $ The commutators between the
operators are
\begin{eqnarray}
\left[\hat{L}_z,\hat{L}_x \right]&=&
\frac{4i}{N}\hat{L}_y,  \\
%-----------------------------------------------
\left[\hat{L}_z,\hat{L}_y \right]&=&
-\frac{4i}{N}\hat{L}_x,   \\
%-----------------------------------------------
\left[\hat{L}_x,\hat{L}_y \right] &=&
\frac{i}{N}\left[\left(1-\hat{L}_z\right)\left(1+3\hat{L}_z\right)+4d^2\right]
\notag \\
&& +\frac{4i}{N^2}\left(1+\hat{L}_z\right).
\label{commu}
\end{eqnarray}
The generalized Bloch surface is determined by the conserved relationship
\begin{eqnarray}
(\hat{L}_x)^2+(\hat{L}_y)^2 =
\left(1+\hat{L}_z+\frac{4}{N}\right)\left[\left(1-\hat{L}_z\right)^2+4d^2\right].
\label{hconserbb}
\end{eqnarray}
Hamiltonian (\ref{heter}) becomes $\hat{H}=-\frac{N}{4} \gamma
\hat{L}_z+\frac{N}{4\sqrt{2}} v \hat{L}_x$. The Heisenberg equations
are $ i \hbar \frac{\textit{d}}{\textit{d}t}\hat{L}_j =
\left[\hat{L}_j, \hat{H}\right], j=x,y,z$. In the mean field limit
where $N\rightarrow\infty$, it is appropriate to replace
 $L_x$, $L_y$,
and $L_z$ by their expected values $u,w$ and $s$, respectively.
Noting the constraint $u^2+w^2=(1+s)\left((1-s)^2-4d^2\right)$ and
introducing the conjugate angular variable $\theta=\arctan( w/u )$
denoting  the relative phase between atoms and molecules, the
Heisenberg equations can be replaced by a classical Hamiltonian of
the form
\begin{eqnarray}
\mathcal{H}^d_m = -\frac{\gamma}{\hbar} s +
\frac{v}{\sqrt{2}\hbar}\sqrt{(1+s)\left[(1-s)^2-4d^2\right]}\cos
\theta, \label{hclas}
\end{eqnarray}
and the canonical equations,
\begin{eqnarray}
d \theta/d t  &=& -\frac{\gamma}{\hbar} - \frac{v}{2\sqrt{2}\hbar}\frac{(1-s)(1+3s)+4d^2}
{\sqrt{(1+s)\left[(1-s)^2-4d^2\right]}} \cos
\theta, \\
d s/d t &=& \frac{v}{\sqrt{2}\hbar}\sqrt{(1+s)\left[(1-s)^2-4d^2\right]} \sin (\theta ), \label{hcannon}
\end{eqnarray}
The  fixed points  of the above system have been obtained by setting
$\dot{s}= \dot{\theta}=0$. The energies for these fixed points make
up energy levels of the system, as shown in Fig.\ref{energy}c,d. The
structure of these energy levels changes dramatically compared to
the homonuclear case. There are always two fixed points
corresponding to two branches of energy levels. Moreover, for a
large population imbalance between two species, for example $d=0.9$,
the energy levels tend to parallel  each other (see.
Fig.\ref{energy}d). Thus, the level space  remains almost  constant
and is slightly dependent on the external field.  We therefore
expect that molecular efficiency is very large in this case.
%=========================================================================
\begin{figure}[t]
  \centering
     \includegraphics[width=0.45\textwidth]{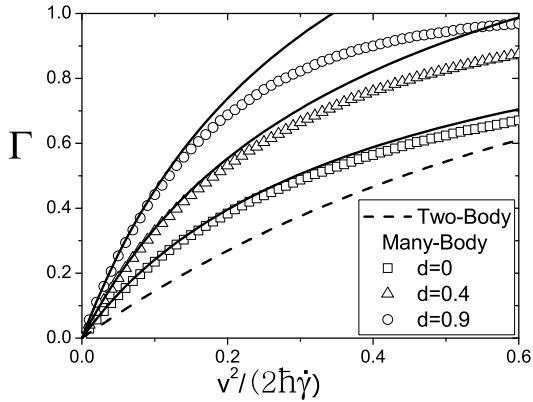}
  \caption{Feshbach heteronuclear molecule production efficiency versus the
  scaled inverse sweep rate. The solid curves are our analytical result.
  For details refer to the text.}
  \label{hete}
\end{figure}
%---------------------------------------------------------------------------
Compared to Hamiltonian (\ref{lzch}), the coupling term in many-body
Hamiltonian (\ref{hclas}) is renormalized by a factor of
$\sqrt{\frac{(1-s)^2-4d^2}{2(1-s)}}$. To derive an approximate
expression, we use
 the  effective coupling $v_{eff}$ as,
$v_{eff}=v\sqrt{\frac{(1-s^*)^2-4d^2}{2(1-s^*)}}$, where  $s^*$ can
be approximately taken as the average between the initial value
$s_0=-1$ and the final value $s_f$, i.e., $s^*=(-1+s_f)/2$. Using
the relation $\Gamma_d=2N_b/N(1-d)=(1+s_f)/2(1-d)$ and formula
(\ref{sf}), we obtain a self-consistent formula for the many-body
conversion efficiency $\Gamma_d$,
\begin{eqnarray}
&&(1-d)\Gamma_{d} \simeq \nonumber\\
 && 1-\exp\left(-\frac{\pi
v^2(2(1-d^2)-\Gamma_d(1-d)(1+d^2))}{4\hbar\dot\gamma}\right).
\end{eqnarray}
The above self-consistent equation for the conversion efficiency
$\Gamma_d$ can be readily solved using the iteration method. The
result is presented in Fig.\ref{hete}. We also numerically solve the
mean field equations for comparison using the Runge-Kutta
step-adaptive algorithm. The agreement is good in the nonadiabatic
regime where the conversion rate is less than 0.6 (see
Fig.\ref{hete}). For $d=0.9$, the regime corresponds to an inverse
scaled sweep rate less than 0.15. It extends to
$v^2/2\hbar\dot\gamma < 0.6$ for $d=0$. Increasing the population
imbalance parameter $d$ means that the effective density of type-2
atoms decreases, but at same time a type-2 atom has more chance to
collide with its  partner, the type-1 atom,  because the density of
type-1 atom increases. The competition between these two effects
leads to an enhancement of heteronuclear molecular conversion
efficiency in the nonadiabatic regime.  Outside the nonadiabatic
regime, our analytic formula overestimates the production
efficiency. This deviation is mainly due to the difference in the
range of $s$ of Hamiltonian (\ref{lzch}) and (\ref{hclas}),
respectively, i.e., it is $[-1,1-2d]$ in the heteronuclear case and
but $[-1,1]$ in the homonuclear case. This complicates the  slowly
sweeping case when we use Eq.(\ref{sf}) as the starting point of our
iteration scheme.

In the nonadiabatic regime, the conversion efficiency of the
heteronuclear molecule can be approximated to
 $\Gamma_d \simeq
\frac{(1+d)8\pi^2\hbar \left|a_{bg} \Delta B\right| n}{ m \dot{B}}$.
Defining the $1/e$ ramp rate $\dot B_{1/e}$ as that at $\dot
B_{1/e}$, $\Gamma_d =1/e$, then, the ratio $n/\dot{B}_{1/e} = \frac{
m }{(1+d)8 e\pi^2\hbar \left|a_{bg} \Delta B\right| }$ is predicted
to be independent of the density but inversely proportional  to the
imbalance parameter.

Experimentally,  our theory may apply to  the $^{85}$Rb-$^{87}$Rb
system. In Ref.\cite{hetero}, the heteronuclear molecules of
$^{85}$Rb-$^{87}$Rb have been produced using the Feshbach resonance
technique, where one BEC and a thermal gas of the second species are
used. The main experimental parameters are $a_{bg}=240a_0$, $\Delta
B=4.9G$, $n=1\times10^{14}cm^{-3}$. We then can calculate that  the
dimensionless inverse sweeping rate $v^2/(2\hbar \dot{\gamma})=0.57$
corresponds to the real sweep rate of a practical magnetic field
$\dot{B}=0.1G/\mu s$. However, to apply our theory, we suggest that
the experiment should be  performed under low atom cloud densities
such as $n\sim 10^{11} cm^{-3}$ with both species prepared as
thermal gas of a few tens of nK.

\section{Conclusion}

In conclusion, we have investigated the dynamics of the Feshbach
formation of the molecules in bosonic systems and show that the
many-body effects greatly modify the picture of two-body molecular
production. With the help of a mean-field classical Hamiltonian, we
derive an analytic expression for the conversion efficiency and
explain the discrepancy between the prediction of the Landau-Zener
formula and the experimental data on $^{85}Rb$. Our theory solves a
puzzle on the formation of Feshbach molecules and gives some
predictions on the formation of heteronuclear diatomic molecules
such as $^{85}$Rb-$^{87}$Rb.

\section*{Acknowledgments}

This work is supported by National Natural Science Foundation of
China (No.10725521,10604009), the National Fundamental Research
Programme of China under Grant No. 2006CB921400, 2007CB814800.
%************************************************************

%************************************************************
\end{document}